\documentclass{aa}
\usepackage[dvips]{graphics}
\usepackage{times}

\begin{document}

\thesaurus{3(13.09.1; 13.18.1; 11.16.1; 12.07.1)}

\title{Infrared imaging of WENSS radio sources
  \thanks{This paper is based on data obtained at the Nordic Optical
     Telescope on La Palma (Canary Islands)}}

\author{D. Villani\inst{1} \and S. di Serego Alighieri\inst{2}}

\institute{Dipartimento di Astronomia e Scienza dello Spazio, L.go
     E.~Fermi 5, I-50125 Firenze, Italy
\and
     Osservatorio Astrofisico di Arcetri, L.go E.~Fermi 5, I-50125
     Firenze, Italy}

\offprints{D. Villani}
\mail{villani@arcetri.astro.it}

\date{Received date / accepted date}        

\maketitle  

\begin{abstract}
   We have performed deep imaging in the IR $J$- and $K$- bands 
   for three sub-samples of
   radio sources extracted from the {\sl Westerbork Northern Sky Survey},
   a large low-frequency radio survey containing Ultra Steep Spectrum
   (USS), Gigahertz Peaked Spectrum (GPS) and Flat Spectrum (FS) sources.
   We present the results of these IR observations, carried out with
   the ARcetri Near Infrared CAmera (ARNICA) at the Nordic Optical
   Telescope (NOT), providing photometric and morphologic information 
   on high redshift radio galaxies and quasars. 
   We find that the radio galaxies contained in our sample do not show 
   the pronounced radio/IR alignment claimed for 3CR sources. 
   IR photometric measurements of the gravitational lens system 1600+434
   are also presented.
\end{abstract}

 \section{Introduction}

   Deep near-IR imaging has an important role in the studies of
   very distant objects, whose redshift moves the rest frame visible
   light into the near-IR. In fact, from $z \geq 1$ the wavelength
   range where normal stars emit most of their energy and have their
   prominent spectral features (4000 $ \leq  \lambda \leq $ 10000 \AA)
   is shifted in the near-IR, between 1 and 2.5 $\mu$m.
   In addition, the stellar light of $z \ge 1$ galaxies observed in the
   near-IR is less likely to be affected by extinction and by an active
   nuclear component (i.e. a dust scattered AGN) than in the optical and
   is not contaminated by thermal dust emission. Therefore
   the near-IR is of fundamental importance in order to
   investigate the stellar component of high redshift galaxies.

   This paper describes the results of IR imaging of sources selected
   from the {\sl Westerbork~Northern~Sky~Survey}
   (WENSS, Rengelink et al. \cite{rengelink}), a new
   large-scale low-frequency radio survey that covers the whole sky for
   $\delta \ge 30$ degree at 325 MHz, and about a quarter of this region
   at 609 MHz, to a limiting flux density of 18 and 15 mJy
   respectively.
   Contrary to previous IR works concentrated mostly on powerful radio
   sources (e.g. Lilly \& Longair \cite{lilly}, Rigler et al. \cite{rigler}, 
   Dunlop \& Peacock \cite{dunlop}), 
   the radio catalogue from which we selected our sources
   extends to low flux densities and therefore to galaxies with a much
   lower level of nuclear activity.

   The use of IR imaging allows one to address three important questions:
   first, the identification of the sources in these sub-samples
   with their optical/IR counterparts. At high redshift ($z \ge 1$) the
   peak of the spectral energy distribution of galaxies is shifted in
   the near IR where their flux density is therefore larger than in the
   optical by orders of magnitude. So it is easier to identify sources,
   which correspond to distant galaxies, in the IR than in the optical.

   Secondly, IR imaging allows the study of the IR ``alignment effect'',
   i.e. the alignment
   of the IR morphology with the radio axis, as a function of 
   the redshift and
   radio power.
   The alignment effect (see McCarthy \cite{mccarthy} for a review) has been 
   discovered in the optical, and
   demonstrates a strong influence of the AGN on the optical morphology
   and luminosity. However, its origin is still matter of debate and its
   extension to the IR is important to understand it.
   Dunlop and Peacock (\cite{dunlop}) have detected a clear IR alignment
   effect in a 3CR sample of radio galaxies in the redshift range 
   $0.8 \leq z \leq 1.3$ (they show the IR-radio and the optical-radio
   alignment histograms for this subset, claiming that the
   alignment effect is just as clear at $K$ as in the optical) in 
   contrast with the conclusions of Rigler et al. (\cite{rigler}) for the same
   sub-sample of objects.
   Therefore it is necessary to study the IR alignment in different
   samples to shed some light on the origin of this discrepancy
   (probably, much of the apparent discrepancy arises from different
   methods of analysis) and on the possible presence of different components
   which dominate in the IR and in the optical, respectively.

   Finally, the IR observations enable the study of the stellar populations 
   of distant galaxies through the analysis of the light emitted 
   by normal stars, 
   in order to determine, modelling their spectral energy
   distributions by means of stellar population synthesis models,
   the age of the oldest stars and therefore provide us with information
   about the epoch of galaxy formation.
   We emphasize that the starlight analysis is easier and more
   productive in the samples at lower radio power, 
   as are the ones selected from the WENSS survey, 
   since the properties of the stellar
   populations are less contaminated by the AGN component.

   In the next paragraph we discuss the selection criteria of our
   samples. Then we report on our observations and immediate results for
   individual sources. We then analyse the radio-IR alignment and discuss
   the results obtained for a gravitational lens system.

 \section{Selection criteria}

   Our work is part of a large collaboration for optical and infrared
   studies of radio sources from the WENSS catalogue.
   For this purpose, three samples of the WENSS catalogue have been
   defined:

 \begin{enumerate}
    \item  Ultra Steep Spectrum (USS) radio sources with a flux density
    $ 100 < S_{325 MHz} < 500 $ mJy and a steep spectral index cut-off of
    $\alpha_{609}^{325} < -1.1$ ($S \propto \nu^{\alpha}$). 
    Optical and infrared follow-up work has been limited to the 95 
    per cent of this sample which has no identification on POSS.
    These objects are powerful tracers of high redshift galaxies.
    
    \item  Gigahertz Peaked Spectrum (GPS) radio sources located in two
    regions of the survey: one at $15^{h} < R.A. < 20^{h}$ and $58^{o}
    < dec. < 75^{o}$, which is called the {\sl mini-survey} region
    (Rengelink et al. \cite{rengelink}), centered on the North Ecliptic Pole,
    and the other at $4^{h}00^{m} < R.A. < 8^{h}30^{m}$ and 
    $58^{o} < dec. < 75^{o}$. These sources have a spectral index
    cut-off of $\alpha_{609}^{325} > 0.0$ with a convex radio spectrum
    peaked at a frequency of about 1 GHz (Snellen et al. \cite{Snellen}). 
    They are compact luminous objects, at intermediate and high redshift,
    which are interesting both as a special class of AGN and as probes
    of galaxy evolution.

    \item   Flat Spectrum (FS) radio sources with an initial selection
    from the Greenbank Surveys (Condon \& Broderick \cite{condon}, Gregory 
    \& Condon \cite{gregory}) at 5 GHz ($S_{5GHz} > 20$ mJy). 
    The coordinate limits
    are $17^{h} < R.A. < 19^{h}$ and $dec. > 55^{o}$ and the
    spectral index cut-off $\alpha_{609}^{325} > -0.5$.
    They are mostly quasars, up to the highest redshifts, along with a useful 
    number of radio-faint BL Lacs.

 \end{enumerate}
   The sub-samples of sources to be observed in the IR were defined from the
   above samples in the following way:

 \begin{enumerate}
    \item USS sources:
    \begin{enumerate}
       \item Sub-sample A with $16^{h} < R.A. < 19^{h}15^{m}$ 
       and $dec. > 55^{o}$
       and spectral index $\alpha_{5000}^{327} < -1.2$. 
       At the time of selection
       optical $R$-band images and spectra were already available.

       \item Sub-sample B with $1^{h} < R.A. < 2^{h}15^{m}$ and $30^{o} <
       dec. < 40^{o}$. For these objects we had also $R$-band images and
       VLA maps.

       \item Sub-sample C with $23^{h} < R.A. < 2^{h}$. It is a high
       flux control sample supplied by Richard Saunders. It was selected
       to have a flux density larger than 0.9 Jy at 365 MHz, to be about
       a factor 5-10 brighter in the radio than the sources in the rest
       of the sample.
       For these objects we had only VLA maps.

       Given they overlap in right ascension range with the GPS 
       and FS sources, we
       observed in the IR only 9 out of the 69 objects in sub-sample A.
       On the other hand, we observed most objects of sub-sample B (20
       out of 30) and all 7 objects of sub-sample C.

    \end{enumerate}
    \item GPS sources in the {\sl mini-survey} region further
    constrained to $R.A. > 16^{h}$, with optical
    identification in the $R$-band. We have imaged in $K$-band all the
    14 objects in this sub-sample. 8 of them were also imaged in the
    $J$-band.

    \item No further selection was applied a priori on the FS sample.
    However given the available observing time, we imaged in $K$-band 13
    out of the 67 sources in the FS sample. These were chosen to cover
    the full range of flux densities at 5 GHz in the sample and to
    contain a fair number (5) of sources not identified on deep CCD
    frames in the $R$-band.

 \end{enumerate}
   
 \section{Observations and reduction}

   The observations were performed in two runs (August and September 1995),
   using the Nordic Optical Telescope (NOT) at La Palma, 
   equipped with the ARcetri Near-Infrared
   CAmera (ARNICA, Lisi et al. \cite{lisi}) and they
   were part of the 1995 International
   Time Programme for the telescopes on the Canary Islands on
   optical/IR follow-up studies of WENSS sources. 

   ARNICA relies on a 256x256 HgCdTe NICMOS array 
   and was used with a scale of 0.55 arcsec/pixel and a 
   field of view of 2.35 arcmin.

   Tables 1, 2, 3 list the observed sources for the USS, GPS, and FS
   samples respectively.
   In the first columns of the Tables 
   we give some information from other authors on the optical
   identifications of the three samples.

\begin{table*} 
\caption[ ]{Sample of USS sources}
\begin{flushleft}
\begin{tabular}{llllllllllll}
\hline
Object & Sub-sample & z & R.A. & Dec. &
Band & Mag. & Apert. & Notes  \\ 
	  &  &  & (Epoch: J2000)  & &  & & arcsec & & \\
\hline
 0114+331 & B &  & 01 16 51.71 & 33 27 53.2 & K & $15.99 \pm 0.24$ & 5.52 \\
 0114+333 & B &  &             &            & K & $ > 17.73$       & 4.39  \\
	  &   &  &             &            & J & $ > 19.99$       & 7.16  \\
 0114+365 & B &  & 01 17 10.02 & 37 15 16.4 & K & $18.20 \pm 0.57$ & 3.88 \\
 0115+350 & B &  &             &            & K & $ > 18.06$       & 3.28 \\
 0115+392 & B &  & 01 17 55.07 & 39 44 31.6 & K & $17.86 \pm 0.48$ & 4.93 \\
 0123+344 & B &  & 01 26 25.94 & 34 57 24.9 & K & $18.35 \pm 0.52$ & 3.28 & uncertain
          ID \\
 0127+350 & B &  & 01 30 01.54 & 35 22 44.3 & K & $16.57 \pm 0.21$ & 6.02 &
          uncertain ID \\
 0129+353 & B &  & 01 32 14.07 & 35 48 16.0 & K & $16.30 \pm 0.03$ & 3.83 & uncertain
	  ID \\
 0131+312 & B &  & 01 33 57.99 & 31 43 15.5 & K & $18.89 \pm 0.84$ & 3.28  \\
	  &   &  &             &            & J & $19.90 \pm 0.40$ & 3.83  \\
 0133+320 & B &  & 01 36 50.44 & 32 22 52.7 & K & $ > 18.59$       & 3.28  \\
	  &   &  &             &            & J & $20.61 \pm 0.31$ & 3.28  \\
 0134+323 & B &  &             &            & K & $ > 17.68$       & 5.52 \\
 0135+311 & B &  & 01 38 06.44 & 31 32 41.4 & K & $ > 18.72$       & 3.28  \\
	  &   &  &             &            & J & $20.74 \pm 0.22$ & 3.28 & uncertain
	  ID \\
 0136+325 & B &  & 01 39 34.05 & 33 11 10.0 & K & $17.30 \pm 0.26$ & 3.88 &
          uncertain ID  \\
 0136+333 & B &  & 01 38 54.23 & 33 53 56.2 & K & $16.77 \pm 0.07$ & 3.88 & uncertain
          ID \\
	  &   &  &             &            & J & $17.83 \pm 0.06$ & 3.88 & uncertain
	  ID \\
 0137+252 & C &  &             &            & K & $ > 19.04$       & 3.28  \\
	  &   &  &             &            & J & $ > 22.09$       & 3.28  \\
 0139+375 & B &  & 01 42 07.11 & 38 05 46.3 & K & $16.54 \pm 0.20$ & 4.39 & uncertain
          ID \\
 0140+323 & B &  & 01 43 43.72 & 32 53 48.3 & K & $18.40 \pm 0.51$ & 3.28 \\
	  &   &  &             &            & J & $19.73 \pm 0.12$ & 3.28 \\
 0143+360 & B &  & 01 46 39.66 & 36 20 29.1 & K & $17.60 \pm 0.24$ & 3.28 & uncertain
          ID \\
	  &   &  &             &            & J & $20.03 \pm 0.08$ & 3.28 & uncertain
	  ID \\
 0148+330 & B &  & 01 51 14.01 & 33 18 00.7 & K & $17.07 \pm 0.16$ & 3.88 & object
          1 \\
	  &   &  & 01 51 13.94 & 33 18 07.5 & K & $17.51 \pm 0.24$ & 3.88 & object
	  2 \\
	  &   &  & 01 51 13.83 & 33 18 11.9 & K & $18.81 \pm 0.45$ & 3.88 & object
	  3  \\
 0202+300 & C &  & 02 02 42.65 & 30 00 39.3 & K & $ > 18.23$ & 3.28  \\
	  &   &  &             &            & J & $19.90 \pm 0.26$ & 3.88 & uncertain
	  ID \\
 0204+392 & B &  & 02 07 46.91 & 39 37 54.5 & K & $18.45 \pm 0.31$ & 3.88 & 
          uncertain ID \\
	  &   &  &             &            & J & $21.11 \pm 0.67$ & 3.88 & 
	  uncertain ID \\
 0214+343 & B &  & 02 17 06.00 & 34 44 30.9 & K & $17.85 \pm 0.11$ & 3.88 \\
 1702+604 & A & 3.23 & 17 03 35.89 & 60 38 48.7 & K & $18.42 \pm 0.32$ & 3.88 \\
	  &   &      &      &            & J & $ > 20.86$ & 3.88  \\
 1720+690 & A &  & 17 20 33.98 & 69 04 50.4 & K & $17.48 \pm 0.14$ & 3.28 \\
	  &   &  &             &            & J & $19.06 \pm 0.12$ & 3.28 \\
 1758+582 & A & 2.9  &            &            & J & $ > 20.75$ & 3.28  \\
 1813+604 & A &  & 18 13 43.60 & 60 48 47.3 & K & $ > 18.24$ & 3.28  \\
	  &   &  &             &            & J & $19.72 \pm 0.16$ & 3.28 \\
 1821+620 & A & 2.81 & 18 22 19.56 & 62 06 07.3 & K & $16.82 \pm 0.27$ & 4.39 \\
 1842+592 & A &  & 18 43 31.37 & 59 32 57.8 & K & $17.96 \pm 0.19$ & 3.28 \\
 1842+691 & A &  &             &            & K & $ > 17.88$ & 3.28 &  \\
 1854+622 & A &  &             &            & K & $ > 18.56$ & 3.28  \\
 1913+672 & A &  & 19 13 15.56 & 67 32 22.0 & K & $17.01 \pm 0.36$ & 3.88 & uncertain
          ID \\
 2320+122 & C &  & 23 20 07.73 & 12 22 05.4 & K & $18.45 \pm 1.36$ & 3.28 \\
 2321+223 & C &  & 23 21 42.36 & 22 37 56.8 & K & $17.67 \pm 0.78$ & 3.88 & uncertain
          ID \\
 2334+154 & C &  & 23 34 58.49 & 15 45 51.9 & K & $17.78 \pm 0.35$ & 4.39 & uncertain
          ID \\
 2334+313 & C &  & 23 34 15.16 & 31 39 30.9 & K & $18.22 \pm 0.24$ & 3.28 & uncertain
          ID \\
 2351+103 & C &  & 23 51 26.79 & 10 34 56.2 & K & $17.35 \pm 0.61$ & 4.39 & east
          component \\
	  &   &  & 23 51 26.57 & 10 34 55.7 & K & $17.82 \pm 0.87$ & 4.39 & west
	  component \\
	  &   &  &             &            & J & $19.08 \pm 0.08$ & 4.39 & east
	  component \\
	  &   &  &             &            & J & $19.80 \pm 0.15$ & 4.39 & west
	  component \\
\hline
\end{tabular}
\end{flushleft}
\end{table*}

\begin{table*}
\caption{Sample of GPS sources}
\begin{flushleft}
\begin{tabular}{lllllllllll}
\hline
Object & z & R & Type${ }^{*}$ & R.A. & Dec. &
Band & Mag. & Apert.  \\
	  &  &  &  & (Epoch: J2000) &  & & & arcsec &  \\
\hline
 1620+640 &      & 23.58 & G  & 16 21 15.22 & 63 59 14.1 & K & $18.62 \pm 0.41$ 
          & 3.28 \\
 1622+663 & 0.20 & 17.15 & G  & 16 23 04.38 & 66 24 01.4 & K & $13.24 \pm 0.04$ 
	  & 7.70 \\
 1642+670 & 1.91 & 17.03 & Q  & 16 42 21.67 & 66 55 48.6 & K & $15.05 \pm 0.03$ 
          & 4.39 \\
 1647+622 & 2.19 & 20.41 & Q  &             &            & K & $ > 17.92 $ 
          & 3.88 \\
 1657+582 &      & 23.30 & FG & 16 58 05.05 & 58 22 02.4 & K & $17.80 \pm 0.11$ 
	  & 3.28 \\
	  &      &       &    &             &            & J & $19.40 \pm 0.15$ 
          & 3.28 \\
 1746+692 & 1.88 & 19.22 & Q  & 17 46 29.92 & 69 20 35.7 & K & $16.51 \pm 0.12$ 
	  & 3.88 \\
 1808+681 &      & 23.27 & FG & 18 08 12.24 & 68 14 11.0 & K & $ > 18.85$ 
          & 3.28 \\
	  &      &       &    &             &            & J & $20.11 \pm 0.28$ 
          & 3.28 \\
 1819+670 & 0.22 & 17.72 & G  & 18 19 44.22 & 67 08 46.4 & K & $14.44 \pm 0.07$ 
          & 6.60 \\
	  &      &       &    &             &            & J & $15.85 \pm 0.03$ 
          & 6.60 \\
 1841+671 & 0.47 & 20.53 & G  & 18 41 03.68 & 67 18 50.2 & K & $16.80 \pm 0.54$ 
          & 4.39 \\
 1942+721 &      & 23.00 & G  & 19 41 26.69 & 72 21 42.3 & J & $20.36 \pm 0.17$ 
          & 3.28 \\
 1945+602 & 2.70 & 20.42 & Q  & 19 46 13.14 & 60 31 38.0 & K & $15.65 \pm 0.06$ 
          & 3.28 \\
          &      &       &    &             &            & J & $17.49 \pm 0.05$ 
          & 3.28 \\
 1946+704 & 0.101& 16.33 & G  & 19 45 53.40 & 70 55 48.9 & K & $13.81 \pm 0.07$ 
          & 7.68 \\
	  &      &       &    &             &            & K & $13.42 \pm 0.30$ 
          & ${23.66}^{**}$ \\
	  &      &       &    &             &            & J & $14.80 \pm 0.04$ 
          & 7.68 \\
	  &      &       &    &             &            & J & $14.23 \pm 0.24$ 
          & ${23.66}^{**}$ \\
 1954+614 &      & 22.18 & G  & 19 54 56.16 & 61 53 57.1 & K & $16.31 \pm 0.07$ 
          & 3.28 \\
	  &      &       &    &             &            & J & $18.11 \pm 0.05$ 
          & 3.28 \\
 1958+615 & 1.81 & 22.40 & Q  & 19 59 30.07 & 62 06 45.1 & K & $18.07 \pm 0.15$ 
          & 3.28 \\
	  &      &       &    &             &            & J & $19.57 \pm 0.18$ 
          & 3.28 \\
\hline
\end{tabular}
\end{flushleft}
 ${ }^{*}${G = galaxy; FG = faint galaxy; Q = quasar}

 ${ }^{**}${These apertures were used for comparison with optical  
 photometry (Snellen, PhD thesis)}
\end{table*}

\begin{table*}
\caption{Sample of FS sources}
\begin{flushleft}
\begin{tabular}{llllllllllll}
\hline
Object & z & R & R.A. & Dec. &
Band & Mag. & Apert. & Notes \\
	  &   &  & (Epoch: J2000)  &  &  &  & arcsec & \\
\hline
 1701+552 &  & $ > 22.50$ &             &            & K & $ > 18.24 $ & 3.28 \\
 1702+637 &  & $ > 22.50$ & 17 03 04.73 & 63 42 33.8 & K & $18.44 \pm 0.50$ & 3.28
          & uncertain ID \\
	  &  &            &             &            & J & $19.86 \pm 0.19$ & 3.88 
          & uncertain ID \\
 1712+651 &  & $ > 22.50$ &             &            & K & $ > 18.21 $ & 3.28 \\
 1744+678 & 3.40 & 18.50      & 17 44 42.10 & 67 50 46.5 & K & $16.19 \pm 0.08$ 
          & 4.39 \\
 1754+676 & 3.60 & 19.80      & 17 54 22.14 & 67 37 35.9 & K & $17.74 \pm 0.43$ 
          & 3.28 \\
 1807+599 &  & $ > 22.50$ & 18 07 55.56 & 60 00 00.2 & K & $17.99 \pm 0.49$ 
          & 4.93 \\
 1816+615 &  & $ > 22.50$ & 18 16 55.27 & 61 34 44.4 & K & $18.39 \pm 0.17$ 
          & 3.28 \\
 1841+568 &  & 19.25      & 18 42 43.90 & 56 56 40.3 & K & $16.53 \pm 0.20$ 
          & 3.88 & uncertain ID \\
 1846+610 &  & $ > 22.50$ & 18 47 24.45 & 61 07 50.2 & K & $17.59 \pm 0.44$ 
          & 3.88 \\
 1849+643 &  & 14.50      & 18 49 40.90 & 64 25 15.1 & K & $12.12 \pm 0.04$ 
          & 20.35 & bright galaxy \\
 1851+608 &  & 21.00      & 18 52 32.06 & 60 53 13.8 & K & $16.97 \pm 0.42$ 
          & 4.93 \\
 1854+605 &  & 21.00      & 18 54 42.24 & 60 34 58.2 & K & $17.27 \pm 0.35$ 
          & 3.88 \\
 1857+710 &  & 22.00      &             &            & K & $ > 17.08 $ & 4.39 \\
\hline
\end{tabular}
\end{flushleft}
\end{table*}

   All sources were observed in $K$- band. Some of USS and GPS sources,
   in particular those which were not clearly detected on the $K$- band
   images after the quick-look reduction at the telescope, were also
   observed in the $J$-band to increase the chances of an infrared
   detection. 
   We used the method developed
   by Tyson (\cite{tyson}) and collaborators 
   for deep CCD imaging, in order to ensure background-limited
   images and avoid the saturation of the brightest sources.
   To this aim, for each object, the total exposure time was broken up 
   into a number of
   shorter exposures, background-limited, between which
   the telescope was moved in a 9 position raster.
   For calibration purposes we observed standard stars, typically four
   per night, from the ARNICA list (Hunt et al. \cite{Hunt}). 
   The seeing was good throughout
   our runs, between 1.0 and 0.5 arcsec and conditions were mostly
   photometric.

   The reduction of the images employed the IRAF procedures available in
   the ARNICA data reduction package.
   In order to eliminate the spatial variations in the detector response,
   arising from camera vignetting and variation in the quantum efficiency
   of the detector, we applied the flat field correction, that is the 
   division into each
   frame of a sky flat frame.
   As a rule, for each of the 9 positions of the telescope, the 
   data reduction procedures
   compute a weighted median of the images of the other 8 positions 
   (or a subset of
   them) to form the sky frame for flat-fielding. 
   However, in the case of $K$-band images, it was necessary to do the 
   background subtraction 
   from
   each frame before flat-fielding by means of normalized differential sky
   flats (Hunt et al. \cite{hunt})
   in order to eliminate the contribution of the
   thermal emission from the telescope that becomes significant for 
   this band.
   Finally, the frames were combined together to produce the final
   source frame, using a new algorithm, invented by Hook and Fruchter
   (\cite{hook}) in order to combine multiple stacks of dithered, 
   undersampled image frames.

 \section{IR identifications and photometry}

   For the identification of the IR counterparts of radio sources 
   we relied on the following informations: the most accurate 
   radio positions available and the optical positions of stars visible
   in our IR frames. 
   Concerning the radio positions, for the sources observed at the VLA
   we used the VLA coordinates of the radio core. In case the radio core
   was not detected we used the WENSS positions of the radio sources and we 
   found it useful to overlay the VLA radio maps, when available, 
   on our IR frames.
   For positioning the radio structure over our IR images, we needed
   astrometric information for our IR images. This was obtained using
   stellar objects in common between our images and $R$-band CCD images
   for which the astrometric work was already done 
   (Rengelink and Snellen, private communications), or, 
   when these were not available,
   using optical sky maps extracted from the 
   digitized sky survey (DSS) plates available at the Space Telescope 
   Science Institute.

   The accuracy of our astrometric work depends on several factors: the
   uncertainties in the optical right ascension and declination
   positions; the errors in the radio positions; the accuracy in the
   radio/IR overlay procedure. The accuracy of the optical positions 
   from the DSS plates is about 1 to 2 arcsec, depending on the distance
   of the object from the center of the Schmidt plate. 
   The positional errors in the VLA radio maps are about 1 arcsec or
   better.
   The WENSS radio coordinates (used for some USS sources lacking VLA
   radio maps) have a 1 sigma error on the position better than 5
   arcsec. For double sources that do not show an obvious radio core component,
   the extent of the radio source introduces an additional indetermination
   in the position of the IR counterpart. Finally, the
   accuracy of our radio/IR overlay procedure is about 0.5
   arcsec. Taking into account all these uncertainties, we have accepted
   as good identifications those for which the distance between the
   radio and IR position is less than 1.5 to 5 arcsec, depending on the
   astrometric information available.

   Tables 1, 2, 3 list the coordinates of the IR counterparts.
   In many cases, our IR detections provided us with an useful tool in
   order to confirm the identification of the fainter optical sources
   and to find the counterparts of the sources still lacking optical
   identifications. Figure 1 shows the $K$- and $J$-band images of the
   sources for which we have an IR
   counterpart, even if uncertain.

   For all the IR counterparts, we measured the
   magnitudes using circular apertures with the minimum diameter which
   includes all the detected object flux. We tested this procedure by making
   several measurements with different aperture diameters. In a few
   cases we adjusted the photometric aperture to correspond to that 
   used for the same object on the optical image by Snellen. The size of
   the apertures and the measured magnitudes, corrected for galactic
   foreground extinction, are listed in Tables 1, 
   2 and 3 for the three samples respectively. The photometric
   accuracy given in Tables 1, 2, 3 has been evaluated from the measured
   fluctuations of the sky background around the source for
   measurement apertures equivalent to that used for source itself. If no
   IR counterpart is detected we give 3$\sigma$ upper limit based on
   the background fluctuations.
   The $J-K$ colours which we obtain for the USS sources and for the galaxies
   in the GPS sample are within the range of those observed for distant
   radio galaxies (e.g. Lilly \& Longair \cite{lilly}).
    
 \section{Comments on individual objects}
 \subsection{Sample of USS sources}

   {\bf 0114+331}. The object is resolved but round in $K$-band and has
   two companions at 4.5 arcsec to the south-east which are not
   included in our photometry.

   \noindent
   {\bf 0115+392}. The IR counterpart is 
   at 3 arcsec to the west of the
   central radio component.

   \noindent
   {\bf 0123+344}. We give as a tentative identification a faint
   object at 4 arcsec to the south-east of the radio source which
   is visible also in the $R$-band.

   \noindent
   {\bf 0127+350}. We give as a tentative identification a compact
   object near the eastern component of the northern radio lobe.

   \noindent
   {\bf 0129+353}. Our identification is at 3.6 arcsec to the
   south-west of the radio position.

   \noindent
   {\bf 0133+320}. The faint object visible in $J$ but not in $K$
   corresponds to the optical identification (Rengelink, private
   communication).

   \noindent
   {\bf 0135+311}. Our tentative identification corresponds to a
   faint object in the $R$-band.
   
   \noindent
   {\bf 0136+325}. We give the magnitude of the object at 3 arcsec
   to the east of the central radio component. It is visible also in the
   $R$-band.

   \noindent
   {\bf 0136+333}. We have measured the object which is at about half
   way between the two radio lobes.

   \noindent
   {\bf 0139+375}. We give the magnitude of the object at 4 arcsec
   to the south of the radio position.

   \noindent
   {\bf 0140+323}. Our identification is at 6 arcsec to the south
   of a brighter galaxy.

   \noindent
   {\bf 0143+360}. Our tentative identification is at 4 arcsec
   to the north of the radio position.

   \noindent
   {\bf 0148+330}. We have measured the three object shown in the
   figure. The most likely but still uncertain identification is the
   object number 3.

   \noindent
   {\bf 0202+300}. We give as a tentative identification the extended
   object visible only in $J$ at 3 arcsec to the north-west of the
   radio position.

   \noindent
   {\bf 1702+604}. The identification is just to the north of the
   southern radio lobe and corresponds to the galaxy visible in the $R$-band,
   for which the redshift has been measured (Bremer, private
   communication).

   \noindent
   {\bf 1913+672}. Our tentative identification is 2.5 arcsec to the
   north-west of the radio position.

   \noindent
   {\bf 2321+223}. Our tentative identification is at 3 arcsec
   to the north of the radio position.

   \noindent
   {\bf 2334+154}. Of the 3 objects to the east of the radio position,
   we give as a tentative identification the closest one.

   \noindent
   {\bf 2334+313}. Our tentative identification is the central
   object of the 3 around the radio position.

   \noindent
   {\bf 2351+103}. The object has two components in both $J$- and
   $K$-bands which we measured separately.

 \subsection{Sample of GPS sources}

   {\bf 1958+615}. The quasar is at 3 arcsec to the north-west of
   a star (Snellen, PhD thesis).

 \subsection{Sample of FS sources}

   {\bf 1702+637}. Our identification is at 4 arcsec to the north
   of the radio position.

   \noindent
   {\bf 1841+568}. Our identification is at 3 arcsec to the south
   of the radio position.

   \noindent
   {\bf 1854+605}. Our tentative identification is 3.6 arcsec to the
   south-west of the radio position.

 \section{IR--radio alignment effect}
   We give in Table 4 and 5 the radio and IR position angles along with the
   differences between them for the USS and GPS sub-samples of radio galaxies.
   In case of USS sources we used the VLA radio maps provided us from
   Rengelink (private communication) in order to measure the radio
   position angles. After selecting the objects with definite 
   identification, we have not found evidence for an IR alignment effect:
   3 objects, although resolved, show an IR round morphology; 
   4 objects are too faint in the
   IR to say if they are elongated; for 5 objects we have measured radio
   to IR position angle differences clearly not biased toward zero.

   These results, although obtained on a small number of objects,
   suggest that the IR alignment in our sample of USS sources is not as
   strong as detected in the optical on powerful distant radio sources
   (e.g. McCarthy \cite{mccarthy}). Furthermore, we are unable to confirm on our
   sample the claim for a precise IR-radio alignment obtained by Dunlop
   and Peacock (\cite{dunlop}) for 3CR sources at $z \sim$ 1. 
   We suggest that the
   lack of alignment in our sample may be connected with its lower radio
   power. It seems in fact plausible that the alignment effect, if it is
   a manifestation of the nuclear activity on the optical/IR morphology,
   would decrease with radio power which is a measure of nuclear
   activity.

   In the case of GPS sources we have defined the radio position angles
   using the analysis of multifrequency VLBI observations carried out 
   by Snellen (PhD thesis) for our sample of faint objects. 
   In all cases, we chose the lowest frequency maps to enhance
   the importance of the extended parts.
   The IR position angles of GPS galaxies are in general not 
   aligned with the radio axis (see Table 5). 
   The only exception (1942+721) could well
   happen by chance.
   Furthermore, 3 objects show an IR round morphology and 1 object is too
   faint in the IR to measure any orientation. The lack of optical-radio
   alignment in GPS sources has also been noticed by Snellen et al. 
   (\cite{snellen}).

\begin{table}
\caption{Position angles in the USS sources.
For those galaxies indicated by a question mark in column 2, we have
only one radio component. The numbers with colon are uncertain.}
\begin{flushleft}
\begin{tabular}{llll}
\hline
Object & Radio P.A. & IR P.A. & $\Delta$ P.A.(Radio-IR)  \\
	  & (\degr) & (\degr) & (\degr)  \\
\hline
0114+331 & no radio map & round image  \\
0114+365 & ? & 65: \\
0115+392 & 167 & 110 & 57  \\
0131+312 & 127: & 90: & 37:  \\
0133+320 & 150 & too faint  \\
0140+323 & 104 & 80: (J) & 24: \\
0204+392 & 26 & too faint  \\
0214+343 & 73: & 140: & 67:  \\
1702+604 & 176 & too faint  \\
1720+690 & ? & round image  \\
1813+604 & ? & round image  \\
1821+620 & ? & 102  \\
1842+592 & 76 & 13 & 63 \\
2320+122 & no radio map & too faint  \\
2351+103 & no radio map & 80  \\
\hline
\end{tabular}
\end{flushleft}
\end{table}

\begin{table}
\caption{Position angles in the GPS sources. The numbers with colon
are uncertain.}
\begin{flushleft}
\begin{tabular}{llll}
\hline
Object & Radio P.A. & IR P.A. & $\Delta$ P.A.(Radio-IR) \\
       & (\degr) & (\degr) & (\degr)   \\
\hline
1620+640 & 169 & round image \\
1622+663 & 61 & 111 & 50 \\
1657+582 & 30 & 170 & 40 \\
1808+681 & 172 & too faint \\
1819+670 & 44 & 160 & 64 \\
1841+671 & 169 & 50: & 61: \\
1942+721 & 62 & 70 & 8 \\
1946+704 & 27 & round image \\
1954+614 & 150 & round image \\
\hline
\end{tabular}
\end{flushleft}
\end{table}

 \section{IR imaging of the gravitational lens system 1600+434}
   One of the main objectives of the WENSS survey was
   to increase the small number of known gravitational lenses by
   mapping flat-spectrum radio sources, in order to provide for the
   first time a sample large enough to do reasonable statistics.
   In addition, optical/IR follow-up studies of confirmed
   lens systems allow the determination of the redshifts of the lensed objects 
   and provide information about the lensing galaxy type. 
   In our run of 1995 August, we performed IR imaging of the object 
   1600+434, a lensed quasar at redshift 1.6, which consists of two
   images, separated by 1.4 arcsec, both in the radio and in the optical
   (Jackson et al. \cite{jackson}).
   The lens has been detected in the $K$- and $J$-bands but not resolved, 
   although the seeing was good, probably because our pixel was 0.55 arcsec 
   and undersampled the point spread function (PSF). 
   In order to estimate the IR magnitudes and colours of the lensing galaxy,
   we have tried to carefully subtract the two quasar images using 
   the astrometry of these two point sources from Jackson et al. 
   (\cite{jackson})
   and the astrometry of the lensing galaxy obtained from HST images 
   (Jackson, private communication). 
   We computed the pixel positions of the two quasar components 
   assuming that the position of the lensing galaxy coincided with the 
   light peak of the image. The quasars subtraction procedure was carried 
   out constructing two empirical two-dimensional PSFs.
   To this aim, after selecting some isolated stars in the field, 
   we calculated the FWHM for all these objects and used its mean value 
   to construct a gaussian PSF.  
   The intensity of the PSF to be subtracted for the North Western image
   of the quasar has been obtained with a trial and error procedure
   which best eliminates the extension to the North West of our blended
   $J$- and $K$-band images. The intensity of the PSF corresponding to
   the other quasar image was then fixed by the flux ratio in the radio
   (1.25) (Jackson et al. \cite{jackson}).
   Figure 2 shows the $J$-band image of the lens system before and
   after subtraction of the two quasar images. This procedure allowed us
   to estimate the magnitude of the lensing galaxy in the two bands,
   which is listed in Table 6 together with the observing parameters.

\begin{table}
\caption{Observing parameters of the gravitational lens system 1600+434}
\begin{flushleft}
\begin{tabular}{lllll}
\hline
Observing date & Obs. time & Band & {Mag.}${}^{*}$ & Apert.  \\
       &  min.   & & & arcsec \\
\hline
14 Aug 1995 & 30 & K & $16.65 \pm 0.12$ & 4.39 \\
16 Aug 1995 & 30 & J & $19.78 \pm 0.12$ & 4.39 \\
\hline
\end{tabular}
\end{flushleft}
${}^{*}${Lensing galaxy only}
\end{table}

   We would like to stress that the photometric information on the
   lensing galaxy which we extract from our data is based on the VLA and
   HST astrometry and on the radio flux ratio for the two quasar images
   and is only partially affected by our relatively poor detector
   sampling. The stated accuracy in the lensing galaxy magnitude takes
   into account the uncertainty in the registration of the positions 
   of the quasar images on the pixel grid.

 \section{Conclusions}
   We presented the results of deep IR imaging for three sub-samples of
   radio sources of the WENSS catalogue: 36 USS, 14 GPS and 13
   FS sources. 
   We show images in the $J$- or $K$-bands of the identified objects:
   29 USS sources, 13 GPS sources and 10 FS sources. 
   We have found IR counterparts for 44\% of USS sources, 93\% of GPS 
   sources and 77\% of FS sources. In addition, there are uncertain IR 
   counterparts for 36\% of USS sources and 15\% of FS sources.
   In some cases, we find the IR counterparts of radio sources still lacking
   optical identifications. We remark that for the USS sources we find a
   larger percentage of certain IR identifications for the A and B sub-samples
   than for the high flux control sub-sample C.

   We analyse the IR-radio alignment effect on 15 USS and 9 GPS radio
   galaxies and we do not find evidence for an alignment of the IR
   morphology with the radio axis.

   Finally, we present the $J$-band image of the
   gravitational lens system 1600+434, before and after subtraction of
   the quasar images. This procedure allow us to provide an estimate of
   the $J$- and $K$-magnitudes of the lensing galaxy.

 \acknowledgements{We thank Ruggero Stanga and Carlo Baffa for their 
 support to the
 observations with ARNICA at the NOT (La Palma). We are grateful to Leslie Hunt
 for her kindness in helping with the data reduction, and to Richard
 Hook for providing the ``drizzling'' software. We thank George Miley for
 the hospitality at the Sterrewacht Observatory of Leiden and 
 Malcolm Bremer, Andrea Cimatti, Roeland Rengelink and Ignas Snellen for 
 useful discussions and for providing unpublished data.}

 \begin{figure*}
 \caption[]{$K$- and $J$-band images (grey scale) of USS, GPS and FS 
   radio sources. The arrows point to the 
   IR counterparts.  The crosses indicate the VLA positions of the 
   radio cores or of the radio centroids; when available, the VLA 
   radio contours are superimposed 
   on the IR images. In the case of 0114+331 we show also contour levels of
   the IR flux, in order to enhance the two companions of the source 
   at 4.5 arcsec to the south-east.
   In a few cases, bad pixels or columns are present, although they were
   corrected for the photometry.}
 \end{figure*}

 \begin{figure*}
 \caption[]{The $J$-band image (North is to the top and East to the left)
   of the gravitational lens system 1600+434, before (a) and after (b)
   the subtraction of the two quasar images.  
   Of the two bright objects in the field the one to the north-west is the
   partial superposition of the two quasar images and of the lensing
   galaxy, while the one to the south-east is a nearby galaxy.}
 \end{figure*}

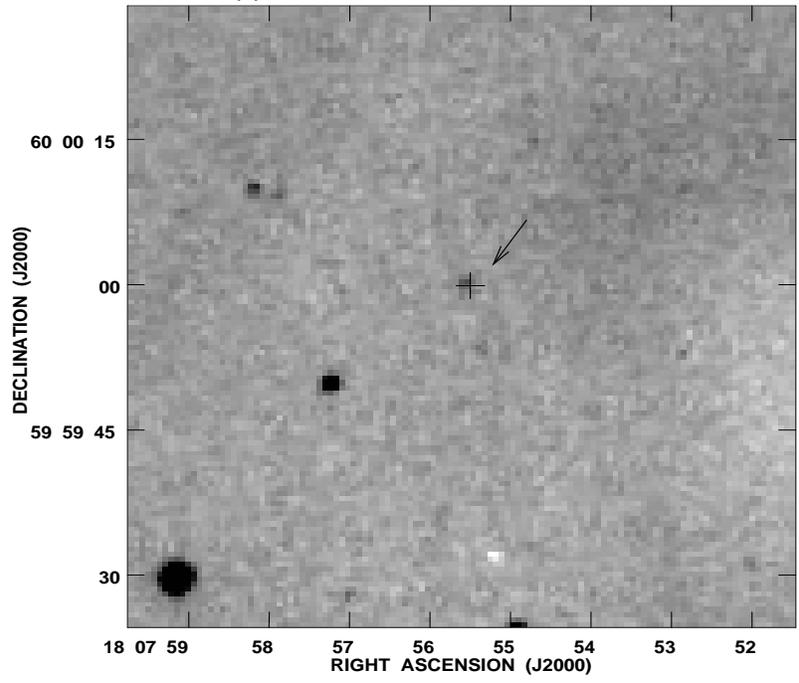
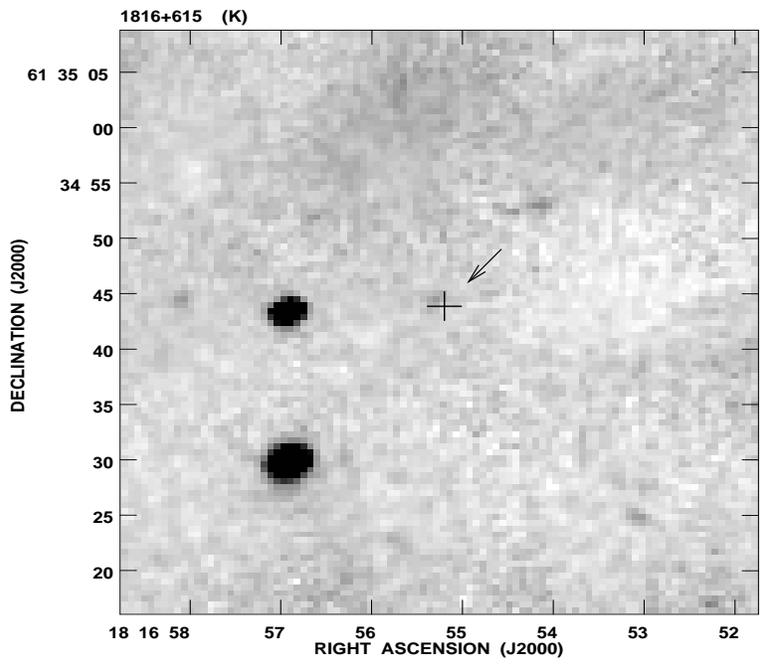
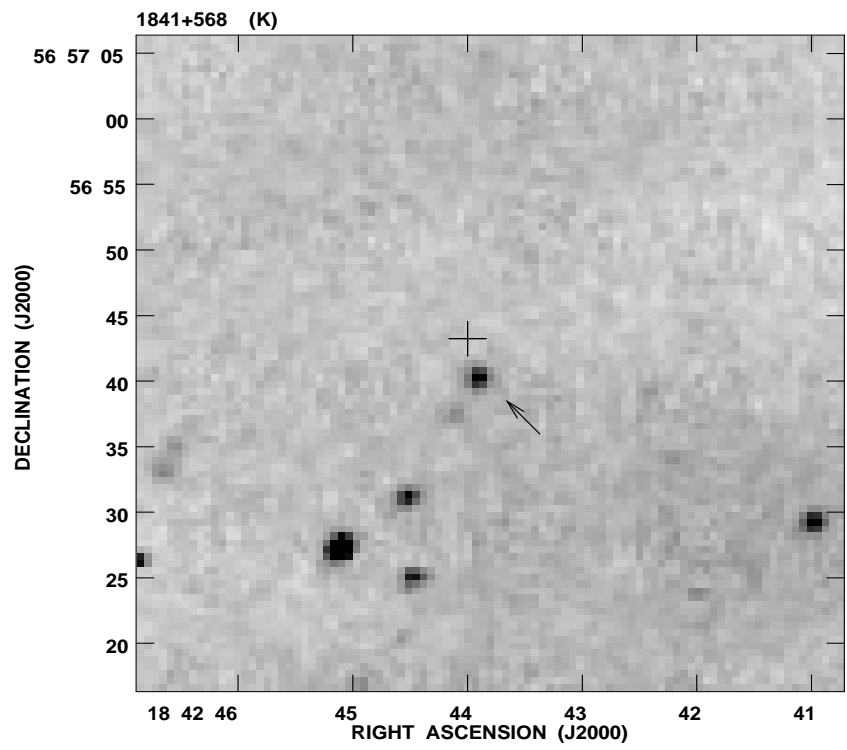
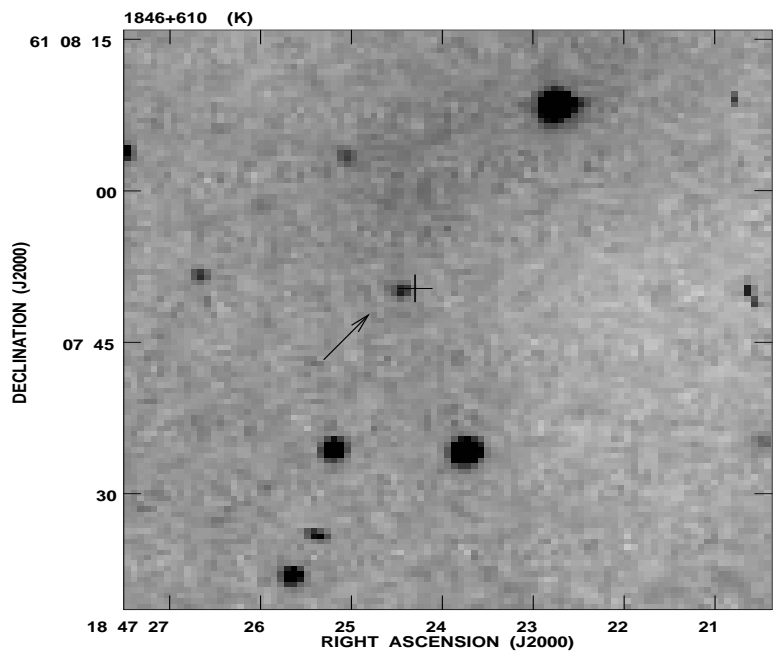
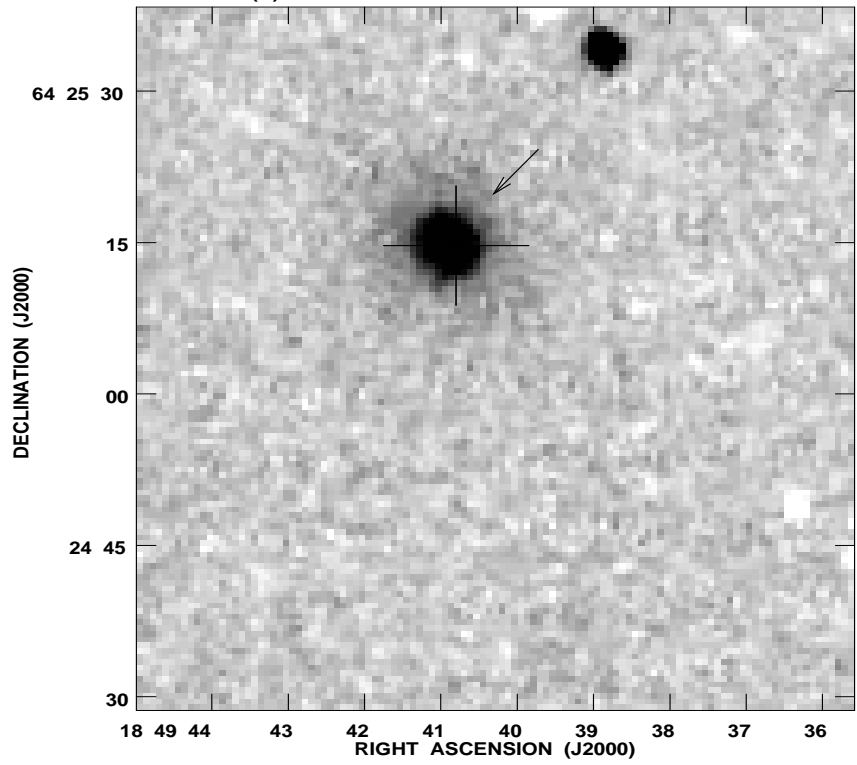
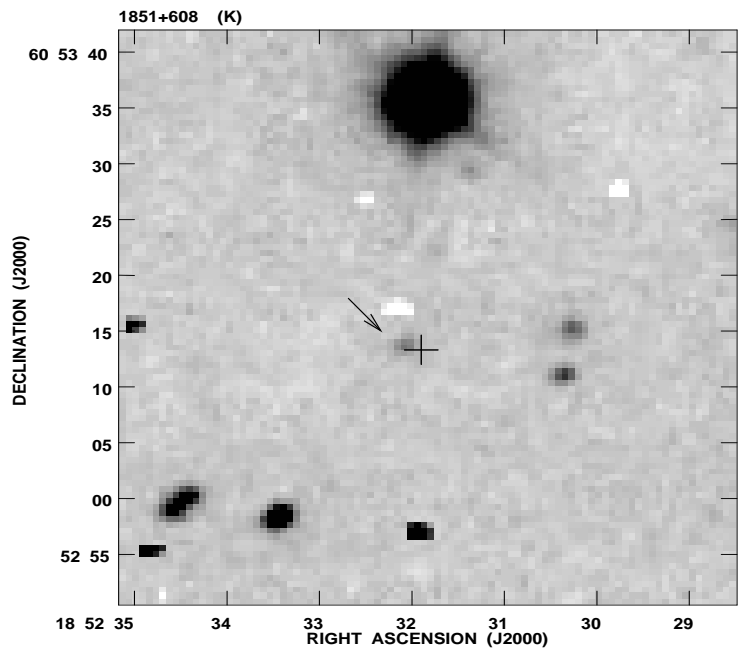
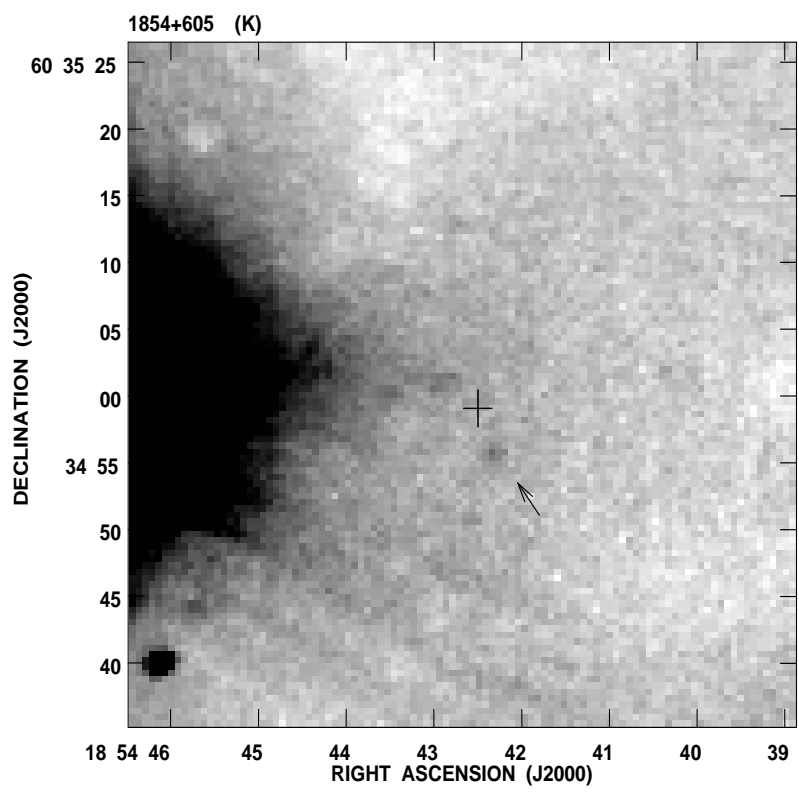
\end{document}